\documentclass[twocolumn,aip,jcp,amsfonts,amsmath,amssymb,floatfix]{revtex4}
\usepackage[version=3]{mhchem}
\usepackage{graphicx}
\usepackage{bm}
\usepackage{mathtools}
\usepackage{mathrsfs}

\begin{document}
\title{Proper and improper zero energy modes in Hartree-Fock theory and their relevance for symmetry breaking and restoration}
\author{Yao Cui}
\thanks{These authors have contributed equally to this research.}
\affiliation{Department of Chemistry, Rice University, Houston, Texas, 77251-1892, USA}
\author{Ireneusz W. Bulik}
\thanks{These authors have contributed equally to this research.}
\affiliation{Department of Chemistry, Rice University, Houston, Texas, 77251-1892, USA}
\author{Thomas Henderson}
\affiliation{Department of Chemistry and Department of Physics and Astronomy, Rice University, Houston, Texas, 77251-1892, USA}
\author{Carlos A. Jim\'enez-Hoyos}
\affiliation{Department of Chemistry, Rice University, Houston, Texas, 77251-1892, USA}
\author{Gustavo E. Scuseria}
\affiliation{Department of Chemistry and Department of Physics and Astronomy, Rice University, Houston, Texas, 77251-1892, USA}
\date{\today}

\begin{abstract}
We study the spectra of the molecular orbital Hessian (stability matrix) and random-phase approximation Hamiltonian of broken-symmetry Hartree-Fock solutions, focusing on zero eigenvalue modes.  After all negative eigenvalues are removed from the Hessian by following their eigenvectors downhill, one is left with only positive and zero eigenvalues.  Zero modes correspond to orbital rotations with no restoring force.  These rotations determine states in the Goldstone manifold, which originates from a spontaneously broken continuous symmetry in the wave function.  Zero modes can be classified as improper or proper according to their different mathematical and physical properties.  Improper modes arise from symmetry breaking and their restoration always lowers the energy.  Proper modes, on the other hand, correspond to degeneracies of the wave function, and their symmetry restoration does not necessarily lower the energy.  We discuss how the RPA Hamiltonian distinguishes between proper and improper modes by doubling the number of zero eigenvalues associated with the latter. Proper modes in the Hessian always appear in pairs which do not double in RPA.  We present several pedagogical cases exemplifying the above statements. The relevance of these results for projected Hartree-Fock methods is also addressed.
\end{abstract}
\maketitle

\section{Introduction
\label{sec:Introduction}}
Although the Hartree-Fock (HF) determinant is understood as variationally minimizing the energy, satisfaction of the self-consistent-field (SCF) equations\cite{roothaan1951new,hall1951molecular,roothaan1960self} guarantees no more than a stationary point in the energy surface.  In other words, the energy is made invariant to any first-order variation of the spinorbitals in the Slater determinant.  To determine whether the energy is a local minimum, a local maximum, or a saddle point, examining the second-order variation is necessary\cite{thouless1960stability,vcivzek1967stability,seeger1977self}.  This problem was first discussed by Overhauser\cite{overhauser1960structure} in 1960 when he found an alternative lower-energy solution of an infinite linear system of fermions. The general conditions for the stability of the HF solutions were originally formulated by Thouless\cite{thouless1960stability} and later developed by
others\cite{ostlund1972complex,fukuda1963stability,adams1962stability,lykos1963discussion,koutecky1967unrestricted,vcivzek1967stability,paldus1969stability,paldus1970stability,paldus1970stability2,vcivzek1971stability}.  

The Hartree-Fock stability test is related to the diagonalization of the molecular orbital (MO) Hessian matrix.  If the Hessian is positive definite, the solution corresponds to at least a local minimum.  If there is a negative eigenvalue, a lower energy solution can be found by proceeding initially in the direction of the corresponding eigenvector.  When the lowest eigenvalue is identically zero, the story is more complicated.

Often (but not always), a lower energy solution found by the stability test breaks a physical symmetry, such as point group or spin.  If the symmetry that has been broken is a continuous one-body symmetry, then the Hessian matrix of the broken symmetry solution will have a zero eigenvalue, and the associated eigenvector will point in the direction required to restore that symmetry\cite{blaizot1985quantum}.  More precisely, it will point towards another determinant in the so--called Goldstone manifold associated with the broken symmetry, where the Goldstone manifold consists of other determinants which have broken the same symmetry.  The various determinants in the Goldstone manifold are degenerate, non-orthogonal (in finite systems), and can be connected by a rotation with a continuous parameter.  Diagonalizing the Hamiltonian in the basis of these determinants restores the symmetry\cite{yoccoz,scuseria2011projected,jimenez2012projected}. If the symmetry that has been broken is instead discrete (such as complex conjugation), then the Hessian matrix of the broken symmetry solution will not have a zero-energy eigenvector pointing in the direction of symmetry restoration.  Nonetheless, a set of degenerate and non-orthogonal states which have broken the same symmetry in different ways can be found and the Hamiltonian can be diagonalized in this basis to restore the symmetry.  Unlike in the case of continuous symmetries, however, these degenerate and non-orthogonal states cannot be reached by a rotation with a continuous parameter.  While the Hessian matrix of a stable but broken-symmetry Hartree-Fock wave function will have zero eigenvalues, the converse is not always true: zero eigenvalues of the Hessian matrix do not necessarily indicate broken-symmetry Hartree-Fock solutions.

We find it convenient to classify the zero-energy eigenvectors of the MO Hessian in terms of modes associated with genuine zero-energy excitations due to degeneracies on the one-hand, and modes which do not correspond to legitimate zero-energy excitations on the other.  We refer to zero-energy excitations as ``legitimate'' if there exists a set of quantum numbers for which degenerate states are expected in the exact solution.  For example, a triplet wave function has a triply-degenerate ground state and therefore two legitimate zero-energy excitations.  Legitimate zero-energy excitations also occur in the thermodynamic limit, where they allow for the appearance of classical behavior.  Artifactual symmetry breaking, on the other hand, arises from approximations. Unrestricted Hartree-Fock, for example, sacrifices good quantum numbers in favor of a variationally superior energy, thereby signalling the appearance of near degeneracies and predicting the failure of symmetry-adapted mean-field methods.  In these cases, projectively restoring the symmetry lowers the energy.

There is a close relation between the Hartree-Fock stability problem and the random phase approximation (RPA)\cite{thouless1960stability}.  An instability in the Hartree-Fock wave function may lead to a mode with imaginary energy in RPA.  When the Hessian has a zero eigenvalue, so too does RPA.   These zero-energy eigenvectors play a crucial role in RPA.  When a Hermitian one-body symmetry is broken, the RPA develops zero eigenvalues which are due not to an intrinsic excitation of the system but to a motion without a restoring force.  The corresponding RPA eigenvector is called a spurious mode.  Nuclear physicists have studied spurious modes associated with different types of symmetry breaking, such as translation, rotation and so on,\cite{blaizot1985quantum} but the quantum chemistry community has paid relatively little attention to the different physical reasons for zero eigenvalues in the Hartree-Fock stability problem \cite{stuber2003symmetry,li2009symmetry,li2009symmetry2}.

In this work, we study the zero-energy eigenvalues and eigenvectors of the general Hartree-Fock spinorbital Hessian and the associated RPA problem, aided by the detailed discussion of the zero-energy modes of quadratic bosonic Hamiltonians given by Colpa in Refs. \onlinecite{colpa1986diagonalization,colpa1986diagonalization2}.  We extend this analysis to draw a more detailed analogy with the symmetries of the mean-field wave function and prove certain matrix-algebraic relations that allow us to shed some light on the zero-energy modes.  To the best of our knowledge, this is the first detailed study of this problem in the quantum chemistry community.  Our motivation for embarking in this study is the relevance of zero energy modes in symmetry breaking and restoration, a subject that has recently received much of our attention\cite{scuseria2011projected,jimenez2012projected}.  We begin in Sec. \ref{sec:Symmetries} by briefly reviewing the symmetries of the electronic structure Hamiltonian before turning to the HF stability conditions in Sec. \ref{sec:HF stability}.  In Sec.\ref{sec:RPA} we recall the equations and properties of RPA.  Section \ref{sec:zero modes} presents mathematical results needed to distinguish between improper modes arising from artifactual symmetry breaking and proper modes corresponding to legitimate zero-energy excitations, as discussed in Sec. \ref{sec:sym}.  Section \ref{sec:dis} discusses benchmark results that we hope will be pedagogical, illustrating our main points.  We close with concluding remarks in Sec. \ref{sec:conclusion}.

\section{Theory
\label{sec:Theory}}
\subsection{Symmetry in Electronic Structure
\label{sec:Symmetries}}
The classification of the symmetries of the electronic Hamiltonian within the Born-Oppenheimer approximation was first carried out by Fukutome \cite{fukutome1981unrestricted} and recently reviewed and extended to include restricted open shells by Stuber and Paldus \cite{stuber2003symmetry}.  It is not our intention to discuss this classification extensively, but merely to note which symmetries we must concern ourselves with.

Typical electronic structure Hamiltonians are invariant to spin rotations, so that $\hat{S}^2$ is a (two-body) symmetry.  In addition to $\hat{S^2}$ symmetry, the Hamiltonian is invariant to spin rotations around all arbitrarily chosen axes (\textit{i.e.} invariant to $\hat{S}_n$ for all axes $n$).  Because the operators $\hat{S}_n$ and $\hat{S}_{n'}$ do not commute with each other, we must pick some axis to privilege; by convention, we take this to be the $z$ axis so that the wave function has good quantum numbers $s$ and $m$ corresponding to the eigenvalues $s(s+1)$ of $\hat{S}^2$ and $m$ corresponding to $\hat{S}_z$.  Hartree-Fock wave functions which respect both these symmetries are known as restricted Hartree-Fock (RHF) or, if open-shell, restricted open-shell Hartree-Fock (ROHF).  A Hartree-Fock determinant which respects $\hat{S}_z$ symmetry but not $\hat{S}^2$ symmetry is known as unrestricted Hartree-Fock (UHF), while a determinant which breaks both and allows each spinorbital to be a linear combination of up-spin and down-spin functions is known as generalized Hartree-Fock (GHF).  Note that any UHF solution can be converted into a GHF solution by a global spin rotation (so that the determinant respects $\hat{S}_n$ symmetry for some axis other than the $z$ axis).  We define ``true''  GHF solutions as determinants which cannot be converted to UHF via global spin rotation, so that there is \textit{no} axis of spin quantization for which the wave function is an eigenfunction.  These spin symmetries are all continuous symmetries in the language above.

Distinct from spin symmetry are complex conjugation and time reversal symmetries, both of which are discrete and antiunitary, meaning that we cannot associate good quantum numbers with them \cite{wigner1960normal}.  Additionally, the Hamiltonian commutes with the number operator so that the electronic wave function has a definite number of particles.  While number symmetry can be broken in the Hartree-Fock-Bogoliubov (HFB) mean-field approximation, it is never broken in Hartree-Fock (by definition).  Moreover HFB and HF coincide for Coulombic systems so we do not consider number symmetry further.

Many molecules, finally, have point group symmetry, which may be either discrete or continuous, depending on the point group involved.  Point group symmetry, like spin symmetry, is unitary, and can be associated with good quantum numbers.

\subsection{Hartree-Fock Stability
\label{sec:HF stability}}
Suppose we have a reference single determinant $|0\rangle$ which we construct from a set of occupied spinorbitals which we label by $i$, $j$, $k$, \ldots.  We will also have virtual spinorbitals labeled by $a$, $b$, $c$, \ldots which, together with the occupied orbitals, form a complete single-particle space.  We can mix the occupied and virtual orbitals via
\begin{equation}
\phi_i \to \phi_i + \sum_a D_{ia} \phi_a
\end{equation}
so that, to second order in $\mathbf{D}$, the resulting determinant $|\Phi\rangle$ can be written as
\begin{equation}
|\Phi\rangle = |0\rangle + \sum_{ia} D_{ia} |{}_i^a\rangle + \frac{1}{2} \sum_{ia,jb} D_{ia} D_{jb} |{}_{ij}^{ab}\rangle + \ldots
\end{equation}
where, for example, $|{}_i^a\rangle = a_a^\dagger \, a_i^{}|0\rangle$.

When $|0\rangle$ satisfies the Hartree-Fock equations, the Hartree-Fock energy $E_0 = \langle 0|\hat{H}|0\rangle$ is invariant to first order in $\mathbf{D}$; in other words, the energy is a stationary point in $\mathbf{D}$.  The nature of this stationary point may be ascertained by considering the second-order corrections to the energy\cite{seeger1977self},
\begin{align}
E_2 &= \sum_{ia,jb} 
\Big[D_{ia}^\star \langle {}^a_i | \hat{H} - E_0 | {}^b_j \rangle D_{jb}^{}
\nonumber
\\
   & \hspace{1cm} + \frac{1}{2} \left(D_{ia}^{} \, D_{jb}^{} \langle 0 | \hat{H} - E_0 | {}^{ab}_{ij} \rangle + \mathit{h.c.}\right)\Big]
\\
   &= \frac{1}{2}
    \begin{pmatrix} \mathbf{D} \hfill \\ \mathbf{D}^\star \hfill \end{pmatrix}^\dagger \, 
    \begin{pmatrix} \mathbf{A} \hfill &  \mathbf{B} \hfill \\  \mathbf{B}^\star\hfill &  \mathbf{A}^\star\hfill \end{pmatrix} \, 
    \begin{pmatrix} \mathbf{D} \hfill \\ \mathbf{D}^\star \hfill \end{pmatrix}.
\nonumber
\end{align}
The matrices $\mathbf{A}$ and $\mathbf{B}$ are
\begin{subequations}
\begin{align}
A_{ia,jb} &= (\epsilon_a - \epsilon_i) \, \delta_{ab} \, \delta_{ij} + \langle aj \| ib \rangle
\\
B_{ia,jb} &= \langle ab \| ij \rangle
\end{align}
\label{eq: AB}
\end{subequations}
where $\epsilon_a$ and $\epsilon_i$ are eigenvalues of the Fock operator and the two-electron integrals such as $\langle ab \| ij \rangle$ are
\begin{subequations}
\begin{align}
\langle ab \| ij \rangle &= \langle ab | ij \rangle - \langle ab | ji \rangle,
\\
\langle ab | ij \rangle &= \int \mathrm{d} \tau_2 \int \mathrm{d} \tau_1 \,\phi_a^\star(1) \phi_b^\star(2) \, \frac{1}{r_{12}} \, \phi_i(1) \, \phi_j(2).
\end{align}
\end{subequations}
Note that $\mathbf{A}$ is Hermitian and $\mathbf{B}$ is symmetric; the Hessian matrix
\begin{equation}
\mathbf{M} = \begin{pmatrix} \mathbf{A} \hfill &  \mathbf{B} \hfill \\  \mathbf{B}^\star\hfill &  \mathbf{A}^\star\hfill \end{pmatrix}
\label{eq:DefHessian}
\end{equation}
is therefore Hermitian.  Note also that we will refer to a Hartree-Fock state as stable if the Hessian is positive semi-definite.

\subsection{The Random Phase Approximation
\label{sec:RPA}}
In the random phase approximation one seeks excitation operators
\begin{equation}
\hat{Q}{}^\dagger_\nu = \sum_{ia} X{}^\nu_{ai} a{}^\dagger_a a_i - \sum_{ia} Y{}^\nu_{ai} a{}^\dagger_i a_a
\end{equation}
whose equation of motion delivers the excitation energies of the system of interest, via
\begin{equation}
[\hat{H},\hat{Q}{}^\dagger_v ] |\mathrm{RPA}\rangle = \omega_\nu \hat{Q}{}^\dagger_v |\mathrm{RPA}\rangle
\end{equation}
where $\omega_\nu = E_\nu - E_0$ is the excitation energy.  If one assumes that the RPA ground state $|\mathrm{RPA}\rangle$ is the HF determinant $|0\rangle$ in what is known as the quasi-boson approximation\cite{blaizot1985quantum,ring2005nuclear}, one obtains the excitation energies and corresponding excitation operators from solving a non-Hermitian eigenvalue problem
\begin{equation}
\bm{\eta} \, \mathbf{M} \, \mathbf{Q} = 
\begin{pmatrix} \mathbf{A} &  \mathbf{B} \\ -\mathbf{B}^\star & -\mathbf{A}^\star \end{pmatrix}
\begin{pmatrix} \mathbf{X} \\ \mathbf{Y} \end{pmatrix}
=
\omega
\begin{pmatrix} \mathbf{X} \\ \mathbf{Y} \end{pmatrix}
\label{eq:RPA}
\end{equation}
where
\begin{equation}
\bm{\eta} = \begin{pmatrix} \bm{1}  &  \bm{0}  \\ \bm{0}  &  -\bm{1} \end{pmatrix}
\end{equation}
and
\begin{equation}
\mathbf{Q} = \begin{pmatrix} \mathbf{X} \\ \mathbf{Y} \end{pmatrix}.
\end{equation}

In the quasi-boson approximation, the RPA problem is fully determined by the Hessian matrix of the underlying HF solution.  In particular, since the eigenvalues of Eqn. \ref{eq:RPA} have the physical meaning of excitation energies, they must be real.  This can be guaranteed if the matrix $\mathbf{M}$ is positive semidefinite, and hence the underlying HF solution is stable.  We emphasize that the opposite is not necessarily true: even if the Hessian $\mathbf{M}$ has negative eigenvalues, the RPA matrix $\bm{\eta} \mathbf{M}$ may have only real eigenvalues.  In other words, real RPA eigenvalues do not guarantee that the Hessian is positive semidefinite\cite{Egido}.  For example, in equilateral H$_3$ the UHF Hessian has a negative eigenvalue pointing toward GHF but the UHF-based RPA has only real eigenvalues.

In the rest of the paper we shall focus only on the zero eigenvalues of $\mathbf{M}$, which connect the stability and RPA problems in an even more subtle way, as clearly all zero-energy eigenvectors of $\mathbf{M}$ are also zero-energy eigenvectors of $\bm{\eta} \mathbf{M}$.

\subsection{Proper and Improper Zero Modes
\label{sec:zero modes}}
The structure of the Hessian $\mathbf{M}$ of Eqn. \ref{eq:DefHessian} means that we can always write its eigenvectors in the form
\begin{equation}
\mathbf{V} = \begin{pmatrix} \mathbf{D}\hfill \\ \mathbf{D}^\star \end{pmatrix}.
\label{Eq:HesVec}
\end{equation}
Indeed, we \textit{must} do so in order to cast the eigenvector as an orbital rotation.  Note that $\mathbf{V}^\dagger \bm{\eta} \mathbf{V} = 0$, a condition which we will refer to as a vanishing $\eta$-norm.  We emphasize for clarity that Hessian eigenvectors can have non-zero $\eta$-norm when not written in the form given in Eqn. \ref{Eq:HesVec}, but they then no longer correspond to search directions for orbital mixing.

Meanwhile, the symplectic character of the RPA matrix $\bm{\eta} \mathbf{M}$ guarantees that if $\mathbf{Q} = \left(\begin{smallmatrix} \mathbf{X} \\ \mathbf{Y}\end{smallmatrix}\right)$ is an eigenvector with eigenvalue $\omega$, then $\mathbf{Q}^\prime = \left(\begin{smallmatrix}\mathbf{Y}^\star \\ \mathbf{X}^\star \end{smallmatrix}\right)$ is an eigenvector with eigenvalue $-\omega^\star$.  Note that this means that $\bm{\eta}\mathbf{M}$ must have an even number of zero eigenvalues whereas no such restriction applies to $\mathbf{M}$ itself.  Thus, while all zero-energy eigenvectors of $\bm{\eta} \mathbf{M}$ are also zero-energy eigenvectors of $\mathbf{M}$ and \textit{vice versa}, the multiplicity of their corresponding eigenvalues may or may not be the same.

We are thus faced with the task of classifying the zero-energy eigenvectors of RPA.  In this, we are guided by the discussion of Colpa \cite{colpa1986diagonalization,colpa1986diagonalization2}.  One can classify the zero-energy eigenvectors of RPA as ``proper'' or ``improper'' modes and can determine the number of each type by studying the spectra of $\mathbf{M}$ and $\bm{\eta}\mathbf{M}$. 

The proper modes $\hat{\mathcal{Q}}^\dagger_{\nu,p}$ are zero-energy eigenvectors of the RPA problem that correspond to legitimate excited states with zero excitation energies.  In particular, there exists a physical state
\begin{equation}
|\nu\rangle = \hat{\mathcal{Q}}^\dagger_{\nu,p} |0\rangle.
\end{equation}
The requirement that we can normalize this state in the quasiboson approximation implies that
\begin{equation}
1 = \langle \nu | \nu \rangle \ = \sum_{ia} \Big(X^{\nu \star}_{ai} X^\nu_{ia} - Y^{\nu \star}_{ai} Y{}^\nu_{ia} ) = \bm{\mathcal{Q}}_\nu^\dagger \, \bm{\eta} \, \bm{\mathcal{Q}}_\nu
\label{Eq:eta-norm}
\end{equation}
where $\bm{\mathcal{Q}}_\nu = \left(\begin{smallmatrix} \mathbf{X}_\nu \\ \mathbf{Y}_\nu\end{smallmatrix}\right)$.  In other words, $\bm{\mathcal{Q}}_\nu$ has an $\eta$-norm of 1.  Proper modes have a counterpart eigenvector $\bm{\mathcal{Q}}_{\nu^\prime} = \left(\begin{smallmatrix} \mathbf{Y}_\nu^\star \\ \mathbf{X}_\nu^\star\end{smallmatrix}\right)$ which is also a zero-energy eigenvector of the RPA matrix but which has an $\eta$-norm of $-1$.  As we prove in Appendix \ref{app:norms}, it is always possible to form a linear combination of proper modes to obtain valid orbital rotations with vanishing $\eta$-norm.  Note that because $\bm{\mathcal{Q}}_\nu$ and $\bm{\mathcal{Q}}_{\nu^\prime}$ are distinct zero-energy eigenvectors of $\bm{\eta} \mathbf{M}$, they are both zero-energy eigenvectors of $\mathbf{M}$ as well, though not necessarily in the form desired.

The situation is drastically different for the improper modes $\hat{\mathcal{Q}}{}^\dagger_{\nu,i}$.  Like the proper modes, it is possible to cast the improper modes as valid orbital rotations.  However, improper modes lack a degenerate counterpart.  Thus, the states $\hat{\mathcal{Q}}{}^\dagger_{\nu,i}|0\rangle$ are not normalizable in the sense of Eqn. \ref{Eq:eta-norm}.  Because the improper mode lacks a degenerate counterpart, it corresponds to a \textit{single} zero eigenvalue of the Hessian matrix even though it leads to \textit{two} zero eigenvalues of the RPA problem.  In other words, an improper mode has one linearly independent eigenvector of $\bm{\eta}\mathbf{M}$ corresponding to an eigenvalue with multiplicity two; the RPA matrix is defective, not diagonalizable.  When a system has multiple improper modes, they are all what we will term $\eta$-orthogonal, which means that we have $\bm{\mathcal{Q}}_i^\dagger \bm{\eta} \bm{\mathcal{Q}}_j = 0$ for all $ij$ pairs.  Note that proper modes, when cast in the form of vectors with vanishing $\eta$-norm so that they are valid orbital rotations, are \textit{not} $\eta$-orthogonal.

These properties allow one to count the number of the proper and improper zero modes simply by investigating the spectrum of the Hessian matrix and the RPA problem.  Denoting the number of proper modes as $2p$ and improper as $i$, the number of zero eigenvalues of $\mathbf{M}$ must be $2p+i$, while there are $2p+2i$ zero eigenvalues of $\bm{\eta} \mathbf{M}$.  Note that the proper modes ``physically'' have a degeneracy of $p$, essentially because the physically meaningful eigenvectors of the RPA problem have positive $\eta$-norm.

For more rigorous discussion of the properties of the improper modes, the reader is referred to Appendix \ref{app:norms} and to Ref. \onlinecite{colpa1986diagonalization}.

\subsection{Improper Modes and Symmetry-Breaking
\label{sec:sym}}
In the following section, we will show that HF wave functions that break continuous symmetries of the underlying Hamiltonian lead to an RPA problem with zero modes. Moreover, the occupied-virtual block of the representation of the generator of the broken symmetry corresponds to the zero eigenvector of the RPA matrix.  The discussion below closely follows Refs. \onlinecite{blaizot1985quantum} and \onlinecite{ring2005nuclear}. 

Suppose that $\hat{P}$ is a Hermitian one-body operator which commutes with the Hamiltonian.  From $\hat{P}$ we can form a unitary operator $\hat{U} = \mathrm{exp}(\mathrm{i} \lambda \hat{P})$, with $\lambda$ a continuous parameter.  Because $\hat{P}$ is a one-body operator, it acts to rotate a reference determinant $|0\rangle$ to another determinant $|\tilde{0}\rangle = \hat{U} |0\rangle$.  Because $\hat{P}$ commutes with the Hamiltonian, the expectation values of $\hat{H}$ with respect to $|0\rangle$ and with respect to $|\tilde{0}\rangle$ are the same.  We thus see that if $|0\rangle$ is a solution of the Hartree-Fock equations, so too is $|\tilde{0}\rangle$.  The manifold of states $|\tilde{0}\rangle$ parameterized by the rotation angle $\lambda$ is what we refer to as the Golstone manifold -- a collection of degenerate, non-orthogonal determinants.

The rotated solution $|\tilde{0}\rangle$ is determined by the rotated one-body density matrix $\tilde{\bm{\gamma}}$,
\begin{align}
\tilde{\gamma}_{ij}
 &= \langle \tilde{0} | a{}^\dagger_j a_i | \tilde{0} \rangle
  = \big(\mathbf{U} \bm{\gamma} \mathbf{U}^\dagger \big)_{ij}
\\
 &= \gamma_{ij}  + \mathrm{i} \lambda [ \mathbf{P} , \bm{\gamma}]_{ij} + \mathcal{O}(\lambda^2) = \gamma_{ij} + \delta \gamma_{ij} + \mathcal{O}(\delta^2 \gamma_{ij}).
\nonumber
\end{align}
Here, $\mathbf{U}$ and $\mathbf{P}$ are the matrix representations of $\hat{U}$ and $\hat{P}$, while $\bm{\gamma}$ is the one-body density matrix associated with the original determinant $|0\rangle$.  The new density matrix $\tilde{\bm{\gamma}}$ satisfies the Hartree-Fock equations,
\begin{equation}
[\mathbf{F}(\tilde{\bm{\gamma}}),\tilde{\bm{\gamma}}] = 0
\end{equation}
where $F_{ij} = \frac{\delta E[\bm{\gamma}]}{\delta \gamma_{ji}}$ is the Fock matrix.  Expanding to first order in $\delta \bm{\gamma}$ (or equivalently in $\lambda$), one arrives at
\begin{equation}
\sum_{ij}[\frac{\partial \mathbf{F}[\bm{\gamma}]}{\partial \gamma_{ij}}\delta \gamma_{ij},\bm{\gamma}] + [\mathbf{F}(\bm{\gamma}),\delta \bm{\gamma}] = \bm{0}.
\end{equation}
The occupied-occupied and virtual-virtual blocks of this equation are trivially satisfied.  From the occupied-virtual and virtual-occupied blocks, one obtains
\begin{equation}
\begin{pmatrix} \mathbf{A} &  \mathbf{B}  \\  -\mathbf{B}^\star & -\mathbf{A}^\star \end{pmatrix}
\begin{pmatrix} \mathbf{P}_{ov}    \\ -\mathbf{P}_{ov}^\star  \end{pmatrix}
= \bm{0}
\label{eq:rpa}
\end{equation}
where $\mathbf{P}_{ov}$ is the occupied-virtual part of $\mathbf{P}$ expressed as a vector with a compound occupied-virtual index.  Thus, for any broken single-particle continuous symmetry there will be a zero-energy eigenvector of the RPA problem.  If the symmetry is not broken, in fact, the foregoing still holds but $\mathbf{P}_{ov}$ identically vanishes.  

Unfortunately, these considerations alone do not allow one to identify the nature of the zero-energy eigenvector.  To do so, one must investigate the $\eta$-orthogonality relations among the broken symmetry generators.  Note also that, depending on the system studied, a given symmetry operator may yield a proper or an improper mode.

In order to illustrate this idea a little better, let us consider UHF determinants.  We cannot consider $\hat{S}^2$ directly in this context because $\hat{S}^2$ is a two-body symmetry, which would lead to zeros in the two-particle two-hole RPA.\cite{toyhama}  However, we can consider $\hat{S}_x$ and $\hat{S}_y$, both of which are symmetries of the Hamiltonian and both of which are broken whenever the UHF is not a spin eigenfunction with $s=0$.  As we show with more detail in Appendix \ref{app:spin}, for a UHF determinant which breaks spin symmetry, $\hat{S}_x$ and $\hat{S}_y$ yield a pair of zero-energy modes in RPA.  When the UHF determinant has $m=0$, these two generators yield two improper modes, but when $m \neq 0$, they instead yield a single proper mode which is then doubled.  Similar arguments have recently been offered in rationalizing the number of Nambu-Goldstone bosons in nonrelativistic theories.\cite{watanabe2012unified,Hidaka}  We should emphasize that of course the wave function cannot be a simultaneous eigenfunction of $\hat{S}_x$, $\hat{S}_y$, and $\hat{S}_z$ except in the trivial case when it is an eigenfunction of $\hat{S}^2$ with $s=0$.  But while picking a preferred spin axis may be physically correct, symmetry breaking it remains.

In the case of a discrete symmetry, even though one could still find a set of degenerate Hartree-Fock states, there is no continuous parameter $\lambda$ which generates a continuum of degenerate states; the foregoing argument then no longer holds.

\subsection{Proper and Improper Modes in Symmetry Restoration}
We shall now attempt to connect the improper modes in the RPA with symmetry restoration.  Given an eigenvector $\mathbf{P}$ of the Hessian matrix with zero eigenvalue associated with an improper mode, one can always
find an associated vector $\mathbf{Q}$ \cite{blaizot1985quantum,ring2005nuclear} defined by
\begin{equation}
\bm{\eta} \, \mathbf{M} \, \mathbf{Q} = -\frac{\mathrm{i}}{\mu} \mathbf{P},
\label{defQ}
\end{equation}
where $\mu > 0$ is a constant which can be completely determined by imposing the normalization conditions
\begin{align}
\mathbf{Q}^\dagger \, \mathbf{M} \, \mathbf{Q} &= \frac{1}{\mu},
\\
\mathbf{Q}^\dagger \, \bm{\eta} \, \mathbf{P} &= \mathrm{i},
\\
\mathbf{Q}^\dagger \, \bm{\eta} \, \mathbf{Q} &= 0.
\end{align}
In its diagonal form, the RPA bosonic Hamiltonian \cite{colpa1986diagonalization,blaizot1985quantum,ring2005nuclear} can then be written in terms of the kinetic energy mode of a free-particle with the form $\hat{\mathcal{P}}^2/(2\mu)$, where
\begin{equation}
\hat{\mathcal{P}} = \begin{pmatrix} \mathbf{b}^\dagger & \mathbf{b} \end{pmatrix} \,  \bm{\eta} \, \begin{pmatrix} \mathbf{P}_{ov} \\ -\mathbf{P}_{ov}^\star \end{pmatrix},
\end{equation}
where $\mathbf{b}$ stands for the set of bosonic annihilation operators.  The kinetic energy (inversely proportional to the mass $\mu$ recovered from Eqn. \ref{defQ}) associated with such a mode has a direct connection with the correlation energy obtained by symmetry-projection schemes.  Indeed, if an approximate symmetry-projection is used instead of a full projection (see, \textit{e.g.}, the Kamlah expansion discussed in detail in Refs. \onlinecite{ring2005nuclear,PhysRevC.2.892,PhysRevC.62.054308}) then the correlation energy is expressed in terms of kinetic energy contributions of the same form as those obtained at the RPA level.

If $\mathbf{P}$ is, on the other hand, a proper mode, an associated vector $\mathbf{Q}$ does not exist\cite{Footnote1}.  This is just a reflection of the fact that proper modes do not render the bosonic Hamiltonian defective, as discussed previously. Hence, at the RPA level, the proper modes are described as true zero-energy excitations rather than as free particles with an associated kinetic energy.

One should bear in mind that, if a mode is proper, that does not mean that full symmetry-restoration of the associated operator will not lower the energy. In the case of spin, if the reference determinant is an eigenfunction of $\hat{S}_z$ (UHF) with eigenvalue different from $0$, then one can always diagonalize the Hamiltonian in the Goldstone manifolds generated by $\hat{S}_x$, $\hat{S}_y$.  A lower energy may be obtained if the original UHF state was not an eigenfunction of the total spin ($\hat{S}^2$) operator.

\section{Results and Discussion
\label{sec:dis}}
In this section, we present some pedagogical examples including both atomic and molecular systems to show the difference between proper and improper modes.  We have counted and analyzed the number of the zero eigenvalues of the Hessian $\mathbf{M}$ ($2p+i$) as well as the RPA matrix $\bm{\eta} \mathbf{M}$ ($2p+2i$) in order to calculate the number of proper ($2p$) and improper ($i$) zero modes according to the method we discussed in Sec.\ref{sec:zero modes}.   We will discuss these examples in three cases.  In Case I, only proper modes appear in the system, in Case II, only improper modes appear, and in Case III, both proper and improper modes appear.  Our results are summarized in Tab. \ref{tab:results}, which shows the expectation values of the three spin components $\hat{S}_x$, $\hat{S}_y$, and $\hat{S}_z$, the type of Hartree-Fock determinant, the number of zero eigenvalues of the Hessian and of the RPA matrix, and the number of proper and improper modes for all systems that we have tested.  For the Hartree-Fock wave functions, ``ROHF" or ``RHF", ``UHF", and ``GHF" respectively mean that we have broken neither $\hat{S}^2$ more $\hat{S}_z$ symmetry, that we have broken $\hat{S}^2$ symmetry but not $\hat{S}_z$ symmetry, and that we have broken both.  Note that in the context of UHF, the terminology of singlet, triplet, \textit{etc.} is somewhat of a misnomer and refers to the $m$ quantum number associated with $\hat{S}_z$ instead of to the $s$ quantum number assoicated with $\hat{S}^2$.  Because the theorems on which we rely are valid for positive semi-definite Hessians $\mathbf{M}$, we will focus on stable Hartree-Fock states.

All calculations have been done using a developmental version\cite{g09} of the \texttt{Gaussian} suite of programs, where the Hessian matrix for complex wave functions has been implemented.  Unless otherwise noted, all calculations in this work are carried out using the cc-pVDZ\cite{dunning1989gaussian} basis set.  All the Hartree-Fock solutions have been tested to be stable in the GHF framework.  That is, a solution of the RHF equations is also a solution of GHF, and we have computed the Hessian for RHF determinants as if they were GHF states instead.  In this way, we have guaranteed that we have not excluded eigenvalues of the Hessian merely on the basis of some symmetry.

\begin{table*}[tbp]
\caption{Expectation values of different spin components ($\hat{S}_x$, $\hat{S}_y$ and $\hat{S}_z$), the type of the HF wave function, the number of the zero eigenvalues of the Hessian matrix ($2p+i$) as well as the RPA matrix ($2p+2i$), and the number of proper ($2p$) and improper ($i$) modes for all the systems that we have tested.  An entry of $\star$ for a spin expectation value indicates a position-dependent real number.
\label{tab:results}}
\begin{center}
\begin{tabular}{cccccccccc}
\hline\hline
System
	&		$\langle \hat{S}_x \rangle$
	&		$\langle \hat{S}_y \rangle$
	&		$\langle \hat{S}_z \rangle$
	&		w.f.
	&		$2p+i$
	&		$2(p+i)$
	&		$2p$
	&		$i$
\\
\hline
\ce{H}  &  0  &  0  &  1/2  &  ROHF  &  2   &  2   &  2   &  0  \\
\ce{B}  &  0  &  0  &  1/2  &  ROHF  &  10  &  10  &  10  &  0  \\
\ce{CH3}&  0  &  0  &  1/2  &   UHF  &  2   &  2   &  2   &  0  \\
\hline
\hline
\ce{H2} &  0  &  0  &  0    &  UHF  &  2  &  4  &  0  &  2  \\
\ce{Be} &  0  &  0  &  0    &  GHF  &  3  &  6  &  0  &  3  \\
\ce{H3} &  0  &  0  &  0    &  GHF  &  3  &  6  &  0  &  3  \\
\hline
\hline
\ce{CH2}(80-90$^{\circ}$)
        &  0  &  0  &  0  &  UHF  &  2  &  4  &  0  &  2  \\
\ce{CH2}(87-100$^{\circ}$)
        &  0  &  0  &  1  &  UHF  &  2  &  2  &  2  &  0  \\
\hline
\ce{CO2}(0.9-1.6\AA,180$^{\circ}$)
        &  0  &  0  &  0  &  RHF  &  0  &  0  &  0  &  0  \\
\ce{CO2}(1.64-1.77\AA,180$^{\circ}$)
        &  0  &  0  &  0  &  GHF  &  3  &  6  &  0  &  3  \\
\ce{CO2}(1.8-2.8\AA,180$^{\circ}$)
        &  0  &  0  &  1  &  UHF  &  3  &  4  &  2  &  1  \\
\ce{CO2}(1.8-2.8\AA,170$^{\circ}$)
        &  0  &  0  &  1  &  UHF  &  2  &  2  &  2  &  0  \\
\hline
\ce{O2}(0.9-1.2\AA) 
        &  0  &  0  &  1  &  UHF  &  2  &  2  &  2  &  0  \\
\ce{O2}(1.3-1.4\AA)
        &  0  &  0  &  1  &  UHF  &  3  &  4  &  2  &  1  \\
\ce{O2}(1.45-1.47\AA)
        &  $\star$  &  0  &  $\star$  &  GHF  &  4  &  6  &  2  &  2 \\
\ce{O2}(1.5-2.8\AA)
        &  0  &  0  &  0  &  UHF  &  3  &  6  &  0  &  3  \\
\ce{O2}(1.9-2.8\AA)
        &  0  &  0  &  2  &  UHF  &  3  &  4  &  2  &  1  \\
\hline\hline
\end{tabular}
\end{center}
\end{table*}

\subsection{Case I \label{sec:proper}} 
Case I includes the systems that only have proper modes. Two atomic systems are under consideration: the hydogen atom and the boron atom.  We also include the methylene radical \ce{CH3}.

\textbf{H atom.} Let us begin by considering the simplest example: the hydrogen atom.  The lowest eigenvalues in the Hessian matrix are zero, which give two proper modes, as shown in Table \ref{tab:results}.  The wave function is a spin eigenfunction and the proper modes just correspond to spin flips.  Recall that only the positive $\eta$-norm eigenvectors have physical significance.  This is easy to say in this simple case, since $\mathbf{B} = \bm{0}$ and the positive and negative $\eta$-norm eigenvectors just correspond to eigenvectors of $\pm \mathbf{A}$, where diagonalizing $\mathbf{A}$ is equivalent to performing configuration interaction with single excitations (which in this case is exact).

This simple example illustrates how a broken symmetry gives rise to a zero eigenvalue in the Hessian.  Despite the fact that HF is an exact theory for the hydrogen atom, the choice of the particular spin quantization axis means that two of the three elements of the SU(2) spin group do not commute with the Hartree-Fock density matrix.  Thus, say, $\hat{S}_x$ and $\hat{S}_y$ acquire non-vanishing occupied-virtual matrix elements, which are the zero-energy eigenvectors of the Hessian.

\textbf{B atom.} The next example that we would like to introduce is the boron atom. We have used the STO-6G basis set in order to ensure that UHF and Restricted Open Shell Hartree-Fock (ROHF) coincide.  Note, however, that in general we exclude ROHF from our considerations as it does not correspond to the usual energy minimization.  

For this system, the Hessian matrix has ten zero eigenvalues.  This is simply because the ground state is six-fold degenerate (because we can occupy any of the three $p$ orbitals, each with either $\uparrow$-spin or $\downarrow$-spin), so that there are five zero-energy excitations, which are then doubled in a manner entirely analogous to what we have seen for the hydrogen atom above.  As with the hydrogen atom, all modes are proper, but where the modes in hydrogen merely change the eigenvalue of $\hat{S}_z$, those in boron may also change the eigenvalue of the orbital angular momentum operator $\hat{L}_z$.

\textbf{\ce{CH3} molecule.} As a final example, we consider the methylene radical at its equilibrium structure, for which the ground state HF solution is a UHF determinant with $\langle \hat{S}_z \rangle = 1/2$.  It therefore breaks $\hat{S}_x$ and $\hat{S}_y$ symmetries and, as we expect, gives rise to two proper modes.  This is unlike our previous atomic examples, in that here we break, in addition, $\hat{S}^2$ symmetry.

\begin{figure}[tbp]
\includegraphics[width=0.45\textwidth]{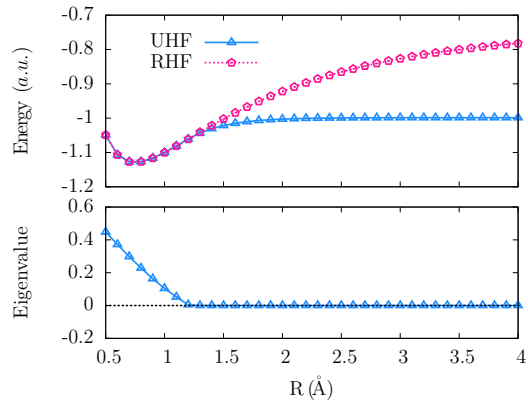}
\caption{Top Panel: Dissociation curves of \ce{H2} computed at the RHF and UHF level. Bottom Panel: The lowest eigenvalue of the Hessian matrix as a function of H--H distance.
\label{fig:h2}}   % H2.eps
\end{figure}

\begin{figure}[tbp]
\includegraphics[width=0.45\textwidth]{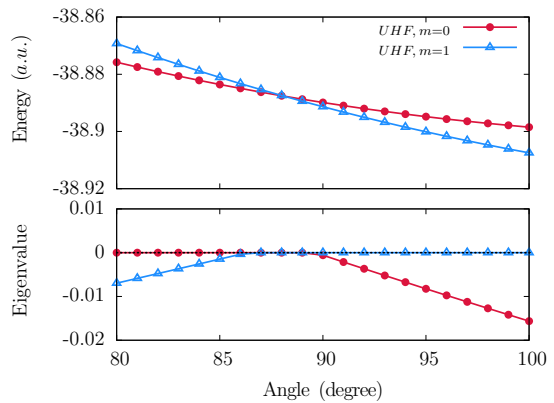}
\caption{Top Panel: Energy versus bond angle of \ce{CH2} computed at the HF level. UHF singlet and triplet energies are shown.  Bottom Panel: The lowest eigenvalue of the Hessian matrix for UHF
singlet and triplet at different bond angles.
\label{fig:ch2}}    % ch2.eps
\end{figure}

\begin{figure}[tbp]
\includegraphics[width=0.45\textwidth]{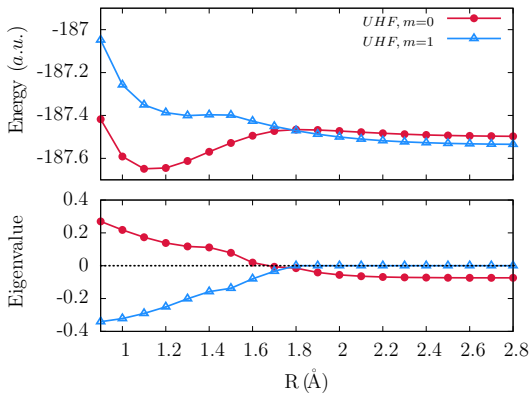}
\caption{Top Panel: Asymmetric dissociation curves of \ce{CO2} computed at the HF level. UHF singlet and triplet are shown.  Bottom Panel: The lowest eigenvalue of the Hessian matrix for UHF singlet and triplet at different distance of \ce{CO2} asymmetric dissociation.  For small bond lengths, the UHF singlet ($m=0$) is the RHF wave function.
\label{fig:co2}}   % co2-asym
\end{figure}

\begin{figure}[tbp]
\includegraphics[width=0.45\textwidth]{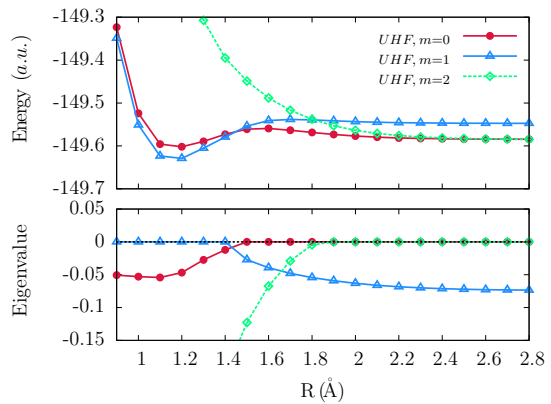}
\caption{Top Panel: Dissociation curves of the \ce{O2} computed at the HF level.  UHF singlet, triplet and quintet solutions are shown.  Bottom Panel: The lowest eigenvalue of the Hessian matrix for UHF singlet, triplet and quintet as a function of O--O distance.
\label{fig:o2}}    % O2
\end{figure}

\subsection{Case II \label{sec:improper}}
Case II includes the systems which only have improper modes.  We show one atomic system (beryllium atom) and two molecular dissociation systems (\ce{H2} and \ce{H3}) in this section. 

\textbf{Dissociation of \ce{H2}.} We begin with our first example of molecular dissociation in the form of \ce{H2}. As shown in Fig. \ref{fig:h2}, before the Coulson-Fischer point\cite{prof1949xxxiv} where the
stable UHF wave function is just RHF, all eigenvalues of the Hessian are positive, as they should be.  Past the Coulson-Fischer point, the UHF and RHF solutions separate.  Table \ref{tab:results} shows that the UHF solution has two zero eigenvalues in the Hessian.  As we have argued earlier, because $m = 0$ in the ground state, both modes are improper.  This is in contrast to the systems in Case I, where $\langle\hat{S}_z\rangle = \tfrac{1}{2}$.

\textbf{Be atom.} The beryllium atom appears to be the simplest atom with a true GHF solution, as originally studied by L\"owdin and Mayer\cite{lowdin1992some}, though this solution appears only in certain basis sets.  Here, we use the STO-6G basis set, for which a stable GHF solution appears.  Since the GHF solution breaks not only $\hat{S}_x$ and $\hat{S}_y$ symmetries but also $\hat{S}_z$ symmetry, this system has three zeros in the Hessian.  Note that in the case of atomic systems, spin symmetry is not the only continuous symmetry of the Hamiltonian.  One might imagine that the orbital angular momentum symmetry may be broken as well, yielding a new set of zero modes.  This is certainly possible in principle, but these modes are not present in this case simply because the occupied-virtual blocks of the matrix representations of the orbital angular momentum operators are linear combinations of those of the spin operators.  This need not be true in all bases, but is true in the minimal basis in which we work.

\textbf{Dissociation of \ce{H3}.} The next example is the dissociation of equilateral \ce{H3}, where we stretch the bond length from 0.9\AA\ to 2.8\AA.  During the whole dissociation process, a true GHF solution is found to be stable with three zeros in the Hessian.  Just as with the beryllium atom, there are three improper modes, and for the same reason.

\subsection{Case III \label{sec:both}} 
In Case III, we consider systems that have both proper and improper zero modes, though not necessarily together at the same geometry.  We focus on the \ce{CH2} molecule, \ce{CO2} asymmetric dissociation and \ce{O2} dissociation.  We should also emphasize that when multiple symmetries are broken simultaneously, or when we have both proper and improper modes simultaneously, it is far from trivial to assign modes to symmetries simply because the diagonalization of the Hessian and the RPA matrix will mix modes together.

\textbf{\ce{CH2} molecule.} We start with the \ce{CH2} molecule, varying the H-C-H angle between 80$^{\circ}$ and 100$^{\circ}$ as shown in Fig. \ref{fig:ch2}.  When the bond angle is less than $\sim$ 87$^\circ$, the UHF singlet is the stable solution.  For bond angles greater than $\sim$ 89$^\circ$ degrees, the triplet is instead the stable solution.  For angles between 87$^\circ$ and 89$^\circ$, both the singlet and triplet display eigenvalues which are numerically zero to the tolerance of our calculation.  Both the singlet and triplet states have two zero modes since the UHF solutions break $\hat{S}_x$ and $\hat{S}_y$ symmetries.  However, these modes are proper for the triplet ground state and improper for the singlet ground state, as we would expect.

\textbf{Asymmetric Dissociation of \ce{CO2}.}  Now, let us turn to the dissociation of \ce{CO2} into an oxygen atom and \ce{CO}, where in the latter the bond length is fixed at 1.16 \AA\ throughout the dissociation process.  Figure \ref{fig:co2} shows that near equilibrium ($r \lesssim$ 1.63 \AA), the RHF is the stable ground state with zeros neither in the Hessian nor the RPA matrix.  Stretching the bond leads to a spin rotation so that the correct dissociation limit can be reached along the $m=1$ surface.  During this spin rotation, a true GHF solution exists (1.64 \AA $\lesssim r \lesssim$ 1.79 \AA).  As in our previous GHF examples, there are three improper modes, corresponding to the three broken spin degrees of freedom.   In the region $r \gtrsim$ 1.80 \AA, the UHF triplet state becomes stable with two proper zero modes and one improper mode.  The two proper zeros correspond to spin, as the UHF has $m=0$.  The improper mode comes from breaking the rotational symmetry of the molecule.  To demonstrate this, we change the O-C-O bond angle to 170$^{\circ}$, eliminating the rotational symmetry of the molecule.  When we do so, the improper mode disappears.

\textbf{Dissociation of \ce{O2}.} Our last example corresponds to the dissociation of the \ce{O2} molecule.  In its ground state, the \ce{O2} molecule is a triplet, dissociating to a pair of triplet atoms.  At the Hartree-Fock level, a triplet can only be described with $m = \pm 1$; the $m = 0$ component of the triplet is a two-determinant wave function.  The result is that UHF cannot describe the dissociation of triplet \ce{O2} to two triplet atoms.  Instead, the proper dissociation limit is reached on the singlet and quintet curves.  We therefore examine singlet, triplet, and quintet UHF states, as shown in Figure \ref{fig:o2}.

As the preceeding discussion implies, both the singlet and quintet curves cross the triplet during the UHF dissociation process, though the singlet has lower energy.  The GHF curve (not shown in Figure \ref{fig:o2}) agrees with the most stable UHF at all points except for a small region (1.45 \AA $\lesssim r \lesssim$ 1.47 \AA) where the UHF singlet and triplet cross and a true GHF solution connects the two via spin rotation\cite{ghf}.  To be more clear, the triplet UHF solution ($m=1$) is stable in the region $r \lesssim 1.4$ \AA, the singlet ($m=0$) for $r \gtrsim 1.5$ \AA, and the quintet ($m=2$) for $r \gtrsim 1.9$ \AA.  Interestingly, as we see in Table \ref{tab:results}, the number of proper and improper zero modes varies from case to case. 

For UHF triplet and quintet stable solutions whose expectation values of $\hat{S}_z$ are 1 and 2, respectively, there are two proper modes coming from spin symmetry breaking.  In contrast, the UHF singlet state has two improper modes arising from spin symmetry breaking, as one would expect.  In the region where a GHF solution exists, we see two proper and one improper modes.  We believe that this is because, unlike with our other GHF examples, one cannot make the expectation value of the spin vector vanish and there must be a preferred direction of spin.

Another observation that can be drawn from Table \ref{tab:results} is that there is another improper mode appearing in all situations where the bond length is larger than 1.2 \AA.  Presumably, this improper mode arises from breaking the rotational symmetry along the O-O bond axis, as happened in \ce{CO2}, although of course we cannot break the D$_{\infty \mathrm{h}}$ symmetry of the molecular Hamiltonian to test this.

\section{Concluding Remarks\label{sec:conclusion}}
While the Hartree-Fock molecular orbital Hessian has a zero eigenvalue whenever a continuous one-body symmetry has been broken, the existence of zero eigenvalues in the Hessian does not imply that symmetry breaking has taken place.  Some zero eigenvalues are associated with genuine degeneracies of the physical Hamiltonian; many others appear to be mere artifacts of the mean-field approximation.  There may be additional zero modes related to the appearance of gauge or quasi-symmetries.\cite{Weinberg}  Zero modes can be classified as proper or improper.  Improper zero modes seem to be related to artifactual symmetry breaking, but individual proper modes may or may not be.  Fortunately, one can distinguish proper from improper modes by counting the number of zero eigenvalues of the related RPA matrix.  Generally, the zero-energy RPA excitation associated with an improper zero mode of the Hessian is artifactual, while the zero-energy RPA excitations associated with proper zero modes of the Hessian may be physically significant.  For example, as we stretch \ce{H2}, the Hessian acquires improper zero modes past the Coulson-Fischer point and thus zero-energy excitations in RPA which are not present in the exact solution of the Schr\"odinger equation.  At dissociation, as discussed in Appendix \ref{app:spin}, the Hessian instead has only proper modes and the zero-energy excitations in RPA become physically meaningful as the true ground state is quadruply degenerate.

When the zero modes are due to symmetry breaking, the symmetry can be projectively restored.  Improper modes are particularly important in this regard, as even an approximate symmetry restoration lowers the energy when the modes are improper.  We believe that the present discussion of the zero modes of the Hessian and the RPA problem will be of interest to the quantum chemistry community, particularly in light of the role the Goldstone manifold plays in projective symmetry restoration\cite{blaizot1985quantum,scuseria2011projected}.

\section{Acknowledgments}
This work was supported by the National Science Foundation (CHE-1102601) and the Welch Foundation (C-0036).

\appendix
\section{Normalization conditions of proper and improper modes
\label{app:norms}}
In the following appendix, we shall prove the $\eta$-orthogonality relations among zero-energy eigenvectors of RPA which allow us to distinguish between proper and improper modes.  The discussion included here is general. We apply this formalism to the spin operators in appendix \ref{app:spin}.  It may be helpful to be familiar with Ref. \onlinecite{colpa1986diagonalization}, which proves the results we rely on here.

We begin by noting that when the Hessian matrix $\mathbf{M}$ of Eqn. \ref{eq:DefHessian} is positive semi-definite, then according to theorem 3.10 of Ref. \onlinecite{colpa1986diagonalization} there exists a matrix $\mathbf{T}$ satisfying $\mathbf{T}^\dagger \bm{\eta} \mathbf{T} = \mathbf{T} \bm{\eta} \mathbf{T}^\dagger = \bm{\eta}$ and $\mathbf{T}^{-1} = \bm{\eta} \mathbf{T}^\dagger \bm{\eta}$, such that
\begin{equation}
\mathbf{T}^\dagger \mathbf{M} \mathbf{T} = \bm{\mathcal{M}} = 
\begin{pmatrix}
\bm{\epsilon}  & \bm{0}           & \bm{0}         & \bm{0}       \\
\bm{0}         & \bm{\alpha}      & \bm{0}         & \bm{\beta}   \\
\bm{0}         & \bm{0}           & \bm{\epsilon}  & \bm{0}       \\
\bm{0}         & \bm{\beta}^\star  & \bm{0}         & \bm{\alpha}^\star  
\end{pmatrix}.
\end{equation}
Here $\bm{\epsilon}$, is real, positive, and diagonal, $\bm{\alpha}$ is Hermitian, $\bm{\beta}$ is symmetric, and the Hermitian submatrix
\begin{equation}
\overline{\bm{\mathcal{M}}} = \begin{pmatrix} \bm{\alpha} & \bm{\beta} \\ \bm{\beta}^\star & \bm{\alpha}^\star \end{pmatrix}
\end{equation}
is such that $\bm{\eta} \overline{\bm{\mathcal{M}}}$ has only zero eigenvalues.

There are several points we should make about this transformation.  First, while $\mathbf{M}$ and $\bm{\mathcal{M}}$ have different eigenvalues, note that if $\mathbf{V}$ is a zero-energy eigevector of $\mathbf{M}$, then $\bm{\mathcal{V}} = \mathbf{T}^{-1} \, \mathbf{V}$ is a zero-energy eigenvector of $\bm{\mathcal{M}}$.  This implies that the zero eigenvalues of $\mathbf{M}$ derive \textit{only} from $\overline{\bm{\mathcal{M}}}$.  This, in turn, means that the number of zero eigenvalues of $\bm{\eta} \mathbf{M}$ is the same as the dimension of $\overline{\bm{\mathcal{M}}}$ and that the zero-energy eigenvectors of $\overline{\bm{\mathcal{M}}}$ can be used to obtain those of $\bm{\eta} \mathbf{M}$.

Let us thus investigate the spectrum of $\overline{\bm{\mathcal{M}}}$.  By Lemma B.4 of Ref. \onlinecite{colpa1986diagonalization}, the (orthonormal) eigenvectors of $\overline{\bm{\mathcal{M}}}$ can be chosen as the union of two disjoint sets $\mathscr{P}$ and $\mathscr{I}$. where
\begin{subequations}
\begin{align}
\mathscr{P} &= \left\{
\begin{pmatrix} \bar{\bm{\mathcal{P}}}_1 \\ \bm{0} \end{pmatrix}, \ldots, 
\begin{pmatrix} \bar{\bm{\mathcal{P}}}_p \\ \bm{0} \end{pmatrix}, 
\begin{pmatrix} \bm{0} \\ \bar{\bm{\mathcal{P}}}_1^\star \end{pmatrix}, \ldots,
\begin{pmatrix} \bm{0} \\ \bar{\bm{\mathcal{P}}}_p^\star \end{pmatrix} \right\},
\\
\mathscr{I} &= \left\{
\begin{pmatrix} \bar{\bm{\mathcal{I}}}_1 \\ \bar{\bm{\mathcal{I}}}_1^\star \end{pmatrix}, \ldots,
\begin{pmatrix} \bar{\bm{\mathcal{I}}}_i \\ \bar{\bm{\mathcal{I}}}_i^\star \end{pmatrix},
\begin{pmatrix} \bar{\bm{\mathcal{I}}}_1 \\ -\bar{\bm{\mathcal{I}}}_1^\star \end{pmatrix}, \ldots,
\begin{pmatrix} \bar{\bm{\mathcal{I}}}_i \\ -\bar{\bm{\mathcal{I}}}_i^\star \end{pmatrix}\right\}
\end{align}
\end{subequations}
where $2p+2i$ is clearly the dimension of $\overline{\bm{\mathcal{M}}}$.  Note that because the eigenvectors of $\overline{\bm{\mathcal{M}}}$ are orthonormal, we have
\begin{subequations}
\begin{align}
\bar{\bm{\mathcal{P}}}_k^\dagger \, \bar{\bm{\mathcal{P}}}_l &= \bar{\bm{\mathcal{I}}}_k^\dagger \, \bar{\bm{\mathcal{I}}}_l = \delta_{kl},
\\
\bar{\bm{\mathcal{P}}}_k^\dagger \, \bar{\bm{\mathcal{I}}}_l &= 0.
\end{align}
\label{PINorm}
\end{subequations}
The important point is that \textit{all} eigenvectors in the set $\mathscr{P}$ and only \textit{half} of the eigenvectors in the set $\mathscr{I}$ are zero-energy eigenvectors of $\overline{\bm{\mathcal{M}}}$.  The remaining eigenvectors in $\mathscr{I}$ have instead positive eigenvalues.  In particular, if $\left(\begin{smallmatrix} \bar{\bm{\mathcal{I}}}_k \\ \pm \bar{\bm{\mathcal{I}}}_k^\star \end{smallmatrix}\right)$ is a zero-energy eigenvector, then  $\left(\begin{smallmatrix} \bar{\bm{\mathcal{I}}}_k \\ \mp \bar{\bm{\mathcal{I}}}_k^\star \end{smallmatrix}\right)$ corresponds to a positive eigenvalue.  It should be clear that the set $\mathscr{P}$ consists of the proper modes and $\mathscr{I}$ of the improper modes.  From now on we will assume that the improper modes are all written as $\left(\begin{smallmatrix} \bar{\bm{\mathcal{I}}}_k \\ \bar{\bm{\mathcal{I}}}_k^\star \end{smallmatrix}\right)$, which we can always arrange by multiplying by $\mathrm{i} = \sqrt{-1}$ as needed.

Having identified eigenvectors of $\overline{\bm{\mathcal{M}}}$, we can read off the corresponding eigenvectors of $\bm{\mathcal{M}}$.  In particular, if $\left(\begin{smallmatrix} \bar{\bm{\mathcal{P}}} \\ \bm{0} \end{smallmatrix}\right)$ is a proper zero-energy eigenvector of $\overline{\bm{\mathcal{M}}}$, then we have
\begin{equation}
\bm{\mathcal{M}} \begin{pmatrix} \bm{0} \\ \bar{\bm{\mathcal{P}}} \\ \bm{0} \\ \bm{0} \end{pmatrix} = \bm{\mathcal{M}} \bm{\mathcal{V}}_p = \bm{0}
\end{equation}
and
\begin{equation}
\bm{\mathcal{M}} \begin{pmatrix} \bm{0} \\ \bm{0} \\ \bm{0} \\ \bar{\bm{\mathcal{P}}}^\star \end{pmatrix} = \bm{\mathcal{M}} \tilde{\bm{\mathcal{V}}}_p = \bm{0}.
\end{equation}
Likewise, if $\left(\begin{smallmatrix} \bar{\bm{\mathcal{I}}} \\ \bar{\bm{\mathcal{I}}}^\star \end{smallmatrix}\right)$ is an improper zero-energy eigenvector of $\overline{\bm{\mathcal{M}}}$, then we have
\begin{equation}
\bm{\mathcal{M}} \begin{pmatrix} \bm{0} \\ \bar{\bm{\mathcal{I}}} \\ \bm{0} \\ \bar{\bm{\mathcal{I}}}^\star \end{pmatrix} = \bm{\mathcal{M}} \bm{\mathcal{V}}_i = \bm{0}.
\end{equation}

All of this means that we can find the proper and improper eigenvectors of $\mathbf{M}$ itself, using the relation $\mathbf{V} = \mathbf{T} \, \bm{\mathcal{V}}$ which as we have already noted holds for the zero-energy eigenvectors.  We can then check the $\eta$-norm of $\mathbf{V}_p = \mathbf{T} \, \bm{\mathcal{V}}_p$, $\tilde{\mathbf{V}}_p = \mathbf{T} \, \tilde{\bm{\mathcal{V}}}_p$, and $\mathbf{V}_i = \mathbf{T} \, \bm{\mathcal{V}}_i$.  Generically, we have 
\begin{equation}
\mathbf{V}^\dagger \, \bm{\eta} \, \mathbf{V} = \bm{\mathcal{V}}^\dagger \, \mathbf{T}^\dagger \, \bm{\eta} \, \mathbf{T} \, \bm{\mathcal{V}} = \bm{\mathcal{V}}^\dagger \, \bm{\eta} \, \bm{\mathcal{V}}
\end{equation}
where we have used $\mathbf{T}^\dagger \, \bm{\eta} \, \mathbf{T} = \bm{\eta}$.  One finds
\begin{subequations}
\begin{align}
\mathbf{V}_p^\dagger \, \bm{\eta} \, \mathbf{V}_p &= \bm{\mathcal{V}}_p^\dagger \, \bm{\eta} \, \bm{\mathcal{V}}_p = \bar{\bm{\mathcal{P}}}^\dagger \, \bar{\bm{\mathcal{P}}} = 1,
\\
\tilde{\mathbf{V}}_p^\dagger \, \bm{\eta} \, \tilde{\mathbf{V}}_p &= \tilde{\bm{\mathcal{V}}}^\dagger \, \bm{\eta} \, \tilde{\bm{\mathcal{V}}} = -\bar{\bm{\mathcal{P}}}^\mathsf{T} \, \bar{\bm{\mathcal{P}}}^\star = -1,
\\
\mathbf{V}_i^\dagger \, \bm{\eta} \, \mathbf{V}_i &= \bm{\mathcal{V}}_i^\dagger \, \bm{\eta} \, \bm{\mathcal{V}}_i = \bar{\bm{\mathcal{I}}}^\dagger \, \bar{\bm{\mathcal{I}}} - \bar{\bm{\mathcal{I}}}^\mathsf{T} \, \bar{\bm{\mathcal{I}}}^\star = 0,
\end{align}
\end{subequations}
where we have used the normalization relations of Eqn. \ref{PINorm}.  Similarly, one can readily show that $\mathbf{V}_p$, $\tilde{\mathbf{V}}_p$, and $\mathbf{V}_i$ are mutually $\eta$-orthogonal:
\begin{equation}
\mathbf{V}_p^\dagger \, \bm{\eta} \, \tilde{\mathbf{V}}_p = \tilde{\mathbf{V}}_p^\dagger \, \bm{\eta} \, \mathbf{V}_i = \mathbf{V}_i^\dagger \, \bm{\eta} \, \mathbf{V}_p = \bm{0}.
\end{equation}
Note that in terms of the block structure for $\bm{\mathcal{M}}$, we would have
\begin{equation}
\bm{\eta} = \begin{pmatrix} \bm{1} & \bm{0} & \bm{0} & \bm{0} \\ \bm{0} & \bm{1} & \bm{0} & \bm{0} \\ \bm{0} & \bm{0} & -\bm{1} & \bm{0} \\ \bm{0} & \bm{0} & \bm{0} & -\bm{1} \end{pmatrix}.
\end{equation}

We therefore see that proper modes can be chosen to have non-zero $\eta$-norm while improper modes cannot.  We can of course create proper modes with zero $\eta$-norm so that they have the correct structure to be an orbital rotation by defining, for example, $\mathbf{W}_{\pm} = \mathbf{V}_p \pm \tilde{\mathbf{V}}_p$, but one can easily show that these modes are then not $\eta$-orthogonal.  From the context of orbital rotations, that is, both proper and improper modes have zero $\eta$-norm but improper modes are $\eta$-orthogonal while proper modes are not.

Note finally that the $\eta$-normalized proper modes appear in pairs $\mathbf{T} \, \bm{\mathcal{V}}_p$ and $\mathbf{T} \, \tilde{\bm{\mathcal{V}}}_p$.

\section{Normalization conditions and the spin-symmetry generators
\label{app:spin}}
As we have shown in Sec. \ref{sec:Theory}, a broken continuous one-body symmetry leads to a zero-energy eigenvector of the Hessian, taking the general form $\mathbf{V} = \left(\begin{smallmatrix} \mathbf{P}_{ov} \\ -\mathbf{P}_{ov}^\star \end{smallmatrix}\right)$ where $\mathbf{P}_{ov}$ is a vector consisting of the occupied-virtual elements of the symmetry generator $\hat{P}$.  We can cast this vector as a valid orbital rotation by multiplying by $\sqrt{-1}$.  As an orbital rotation, this vector has zero $\eta$-norm.  It thus follows from our discussion in Appendix \ref{app:norms} that if $\mathbf{V}$ is to correspond to a proper mode, there must be a second zero-energy eigenvector $\tilde{\mathbf{V}}$ which also has zero $\eta$-norm but which is not $\eta$-orthogonal to $\mathbf{V}$.

In this section, we will apply this formalism to the special case of spin symmetry breaking.  We will assume for simplicity that the only zero modes are due to spin symmetry breaking.

Suppose, then, that the underlying Hartree-Fock state is UHF in nature, so that it has broken $\hat{S}^2$, $\hat{S}_x$, and $\hat{S}_y$ symmetry but not $\hat{S}_z$ symmetry.  We can define eigenvectors correspond to $\hat{S}_x$ and $\hat{S}_y$ symmetry breaking:
\begin{subequations}
\begin{align}
\bm{\mathcal{S}}_x &= \begin{pmatrix} \mathbf{S}^x_{ov} \\ -(\mathbf{S}^x_{ov})^\star \end{pmatrix},
\\
\bm{\mathcal{S}}_y &= \begin{pmatrix} \mathbf{S}^y_{ov} \\ -(\mathbf{S}^y_{ov})^\star \end{pmatrix}.
\end{align}
\end{subequations}
Because these modes have zero $\eta$-norm, we can test whether they are proper or improper by simply checking their $\eta$-orthogonality.  One finds
\begin{subequations}
\begin{align}
\bm{\mathcal{S}}_x^\dagger \bm{\eta} \bm{\mathcal{S}}_y
 &= \sum_{ia} \left[\left(\mathbf{S}^x_{ia}\right)^\star \mathbf{S}^y_{ia} - \mathbf{S}^x_{ia} \left(\mathbf{S}^y_{ia}\right)^\star \right]
\\
 &= \sum_{ia} \left[\mathbf{S}^x_{ai} \mathbf{S}^y_{ia} - \mathbf{S}^x_{ia} \mathbf{S}^y_{ai}\right]
\\
 &= \mathrm{Tr}\left(\mathbf{S}^y_{ov} \mathbf{S}^x_{vo} - \mathbf{S}^x_{ov} \mathbf{S}^y_{vo}\right)
\end{align}
\end{subequations}
where in the last line we have converted $\mathbf{S}^x$ and $\mathbf{S}^y$ from vectors to matrices.  Using the cyclic properties of the trace, we can add and subtract $\mathrm{Tr}(\mathbf{S}^y_{oo} \mathbf{S}^x_{oo})$ to get
\begin{equation}
\bm{\mathcal{S}}_x^\dagger \bm{\eta} \bm{\mathcal{S}}_y = \mathrm{Tr}\left(\mathbf{S}^y_{ov} \mathbf{S}^x_{vo} - \mathbf{S}^x_{ov} \mathbf{S}^y_{vo} + \mathbf{S}^y_{oo} \mathbf{S}^x_{oo} - \mathbf{S}^x_{oo} \mathbf{S}^y_{oo}\right).
\end{equation}
This we recognize as
\begin{subequations}
\begin{align}
\bm{\mathcal{S}}_x^\dagger \bm{\eta} \bm{\mathcal{S}}_y
  &= \mathrm{Tr} \left([\mathbf{S}^y,\mathbf{S}^x]_{oo}\right)
\\
  &= \langle [\hat{S}_y,\hat{S}_x] \rangle
\\
  &= -\mathrm{i} \, \langle \hat{S}_z \rangle.
\end{align}
\end{subequations}

Thus, when $\langle \hat{S}_z \rangle = 0$, the modes corresponding to breaking $\hat{S}_x$ and $\hat{S}_y$ in UHF are improper.  Otherwise, they are proper.  Similarly, for a GHF reference, when $\langle \hat{S}_x \rangle = \langle \hat{S}_y \rangle = \langle \hat{S}_z \rangle = 0$, all three modes are improper; otherwise, one is improper and two are proper.

Finally, note that the analysis above holds only when the Hessian has only two zero eigenvalues for UHF or three for GHF.  The improper modes become proper if the Hessian has additional zero modes which are not $\eta$-orthogonal to them; the proper modes, of course, remain proper.  This is the case for the dissociation of H$_2$, for example.  As we have already seen, the Hessian has two improper modes (so the RPA matrix has four zero eigenvalues).  As the bond stretches, the next lowest two eigenvalues of the Hessian approach zero, and at dissociation both the Hessian and the RPA matrix have four zero eigenvalues, indicating that they correspond to proper modes.  This is as we would expect, since each of the two non-interacting hydrogen atoms has two proper modes.


\begin{thebibliography}{41}
\expandafter\ifx\csname natexlab\endcsname\relax\def\natexlab#1{#1}\fi
\expandafter\ifx\csname bibnamefont\endcsname\relax
  \def\bibnamefont#1{#1}\fi
\expandafter\ifx\csname bibfnamefont\endcsname\relax
  \def\bibfnamefont#1{#1}\fi
\expandafter\ifx\csname citenamefont\endcsname\relax
  \def\citenamefont#1{#1}\fi
\expandafter\ifx\csname url\endcsname\relax
  \def\url#1{\texttt{#1}}\fi
\expandafter\ifx\csname urlprefix\endcsname\relax\def\urlprefix{URL }\fi
\providecommand{\bibinfo}[2]{#2}
\providecommand{\eprint}[2][]{\url{#2}}

\bibitem[{\citenamefont{Roothaan}(1951)}]{roothaan1951new}
\bibinfo{author}{\bibfnamefont{C.~C.~J.} \bibnamefont{Roothaan}},
  \bibinfo{journal}{Rev. Mod. Phys.} \textbf{\bibinfo{volume}{23}},
  \bibinfo{pages}{69} (\bibinfo{year}{1951}).

\bibitem[{\citenamefont{Hall}(1951)}]{hall1951molecular}
\bibinfo{author}{\bibfnamefont{G.}~\bibnamefont{Hall}}, \bibinfo{journal}{Proc.
  Roy. Soc. A} \textbf{\bibinfo{volume}{205}}, \bibinfo{pages}{541}
  (\bibinfo{year}{1951}).

\bibitem[{\citenamefont{Roothaan}(1960)}]{roothaan1960self}
\bibinfo{author}{\bibfnamefont{C.}~\bibnamefont{Roothaan}},
  \bibinfo{journal}{Rev. Mod. Phys.} \textbf{\bibinfo{volume}{32}},
  \bibinfo{pages}{179} (\bibinfo{year}{1960}).

\bibitem[{\citenamefont{Thouless}(1960)}]{thouless1960stability}
\bibinfo{author}{\bibfnamefont{D.}~\bibnamefont{Thouless}},
  \bibinfo{journal}{Nucl. Phys.} \textbf{\bibinfo{volume}{21}},
  \bibinfo{pages}{225} (\bibinfo{year}{1960}).

\bibitem[{\citenamefont{{\v{C}}{\'\i}{\v{z}}ek and
  Paldus}(1967)}]{vcivzek1967stability}
\bibinfo{author}{\bibfnamefont{J.}~\bibnamefont{{\v{C}}{\'\i}{\v{z}}ek}}
  \bibnamefont{and} \bibinfo{author}{\bibfnamefont{J.}~\bibnamefont{Paldus}},
  \bibinfo{journal}{J. Chem. Phys.} \textbf{\bibinfo{volume}{47}},
  \bibinfo{pages}{3976} (\bibinfo{year}{1967}).

\bibitem[{\citenamefont{Seeger and Pople}(1977)}]{seeger1977self}
\bibinfo{author}{\bibfnamefont{R.}~\bibnamefont{Seeger}} \bibnamefont{and}
  \bibinfo{author}{\bibfnamefont{J.~A.} \bibnamefont{Pople}},
  \bibinfo{journal}{J. Chem. Phys} \textbf{\bibinfo{volume}{66}},
  \bibinfo{pages}{3045} (\bibinfo{year}{1977}).

\bibitem[{\citenamefont{Overhauser}(1960)}]{overhauser1960structure}
\bibinfo{author}{\bibfnamefont{A.}~\bibnamefont{Overhauser}},
  \bibinfo{journal}{Phys. Rev. Lett.} \textbf{\bibinfo{volume}{4}},
  \bibinfo{pages}{415} (\bibinfo{year}{1960}).

\bibitem[{\citenamefont{Ostlund}(1972)}]{ostlund1972complex}
\bibinfo{author}{\bibfnamefont{N.~S.} \bibnamefont{Ostlund}},
  \bibinfo{journal}{J. Chem. Phys.} \textbf{\bibinfo{volume}{57}},
  \bibinfo{pages}{2994} (\bibinfo{year}{1972}).

\bibitem[{\citenamefont{Fukuda}(1963)}]{fukuda1963stability}
\bibinfo{author}{\bibfnamefont{N.}~\bibnamefont{Fukuda}},
  \bibinfo{journal}{Nucl. Phys.} \textbf{\bibinfo{volume}{44}},
  \bibinfo{pages}{553} (\bibinfo{year}{1963}).

\bibitem[{\citenamefont{Adams}(1962)}]{adams1962stability}
\bibinfo{author}{\bibfnamefont{W.~H.} \bibnamefont{Adams}},
  \bibinfo{journal}{Phys. Rev.} \textbf{\bibinfo{volume}{127}},
  \bibinfo{pages}{1650} (\bibinfo{year}{1962}).

\bibitem[{\citenamefont{Lykos and Pratt}(1963)}]{lykos1963discussion}
\bibinfo{author}{\bibfnamefont{P.}~\bibnamefont{Lykos}} \bibnamefont{and}
  \bibinfo{author}{\bibfnamefont{G.}~\bibnamefont{Pratt}},
  \bibinfo{journal}{Rev. Mod. Phys.} \textbf{\bibinfo{volume}{35}},
  \bibinfo{pages}{496} (\bibinfo{year}{1963}).

\bibitem[{\citenamefont{Kouteck{\`y}}(1967)}]{koutecky1967unrestricted}
\bibinfo{author}{\bibfnamefont{J.}~\bibnamefont{Kouteck{\`y}}},
  \bibinfo{journal}{J. Chem. Phys.} \textbf{\bibinfo{volume}{46}},
  \bibinfo{pages}{2443} (\bibinfo{year}{1967}).

\bibitem[{\citenamefont{Paldus and \ifmmode \check{C}\else
  \v{C}\fi{}i\ifmmode~\check{z}\else \v{z}\fi{}ek}(1969)}]{paldus1969stability}
\bibinfo{author}{\bibfnamefont{J.}~\bibnamefont{Paldus}} \bibnamefont{and}
  \bibinfo{author}{\bibfnamefont{J.}~\bibnamefont{\ifmmode \check{C}\else
  \v{C}\fi{}i\ifmmode~\check{z}\else \v{z}\fi{}ek}}, \bibinfo{journal}{Chem.
  Phys. Lett.} \textbf{\bibinfo{volume}{3}}, \bibinfo{pages}{1}
  (\bibinfo{year}{1969}).

\bibitem[{\citenamefont{Paldus and
  {\v{C}}{\'\i}{\v{z}}ek}(1970)}]{paldus1970stability}
\bibinfo{author}{\bibfnamefont{J.}~\bibnamefont{Paldus}} \bibnamefont{and}
  \bibinfo{author}{\bibfnamefont{J.}~\bibnamefont{{\v{C}}{\'\i}{\v{z}}ek}},
  \bibinfo{journal}{Phys. Rev. A} \textbf{\bibinfo{volume}{2}},
  \bibinfo{pages}{2268} (\bibinfo{year}{1970}).

\bibitem[{\citenamefont{Paldus and \ifmmode \check{C}\else
  \v{C}\fi{}i\ifmmode~\check{z}\else
  \v{z}\fi{}ek}(1970)}]{paldus1970stability2}
\bibinfo{author}{\bibfnamefont{J.}~\bibnamefont{Paldus}} \bibnamefont{and}
  \bibinfo{author}{\bibfnamefont{J.}~\bibnamefont{\ifmmode \check{C}\else
  \v{C}\fi{}i\ifmmode~\check{z}\else \v{z}\fi{}ek}}, \bibinfo{journal}{J. Chem.
  Phys.} \textbf{\bibinfo{volume}{52}}, \bibinfo{pages}{2919}
  (\bibinfo{year}{1970}).

\bibitem[{\citenamefont{{\v{C}}{\'\i}{\v{z}}ek and
  Paldus}(1971)}]{vcivzek1971stability}
\bibinfo{author}{\bibfnamefont{J.}~\bibnamefont{{\v{C}}{\'\i}{\v{z}}ek}}
  \bibnamefont{and} \bibinfo{author}{\bibfnamefont{J.}~\bibnamefont{Paldus}},
  \bibinfo{journal}{Phys. Rev. A} \textbf{\bibinfo{volume}{3}},
  \bibinfo{pages}{525} (\bibinfo{year}{1971}).

\bibitem[{\citenamefont{Blaizot and Ripka}(1985)}]{blaizot1985quantum}
\bibinfo{author}{\bibfnamefont{J.-P.} \bibnamefont{Blaizot}} \bibnamefont{and}
  \bibinfo{author}{\bibfnamefont{G.}~\bibnamefont{Ripka}},
  \emph{\bibinfo{title}{Quantum theory of finite systems}}
  (\bibinfo{publisher}{The MIT Press}, \bibinfo{address}{Cambridge, MA},
  \bibinfo{year}{1985}).

\bibitem[{\citenamefont{Peierls and Yoccoz}(1957)}]{yoccoz}
\bibinfo{author}{\bibfnamefont{R.~E.} \bibnamefont{Peierls}} \bibnamefont{and}
  \bibinfo{author}{\bibfnamefont{J.}~\bibnamefont{Yoccoz}},
  \bibinfo{journal}{Proc. Phys. Soc. A} \textbf{\bibinfo{volume}{70}},
  \bibinfo{pages}{381} (\bibinfo{year}{1957}).

\bibitem[{\citenamefont{Scuseria et~al.}(2011)\citenamefont{Scuseria,
  Jim{\'e}nez-Hoyos, Henderson, Samanta, and Ellis}}]{scuseria2011projected}
\bibinfo{author}{\bibfnamefont{G.~E.} \bibnamefont{Scuseria}},
  \bibinfo{author}{\bibfnamefont{C.~A.} \bibnamefont{Jim{\'e}nez-Hoyos}},
  \bibinfo{author}{\bibfnamefont{T.~M.} \bibnamefont{Henderson}},
  \bibinfo{author}{\bibfnamefont{K.}~\bibnamefont{Samanta}}, \bibnamefont{and}
  \bibinfo{author}{\bibfnamefont{J.~K.} \bibnamefont{Ellis}},
  \bibinfo{journal}{J. Chem. Phys.} \textbf{\bibinfo{volume}{135}},
  \bibinfo{pages}{124108} (\bibinfo{year}{2011}).

\bibitem[{\citenamefont{Jim{\'e}nez-Hoyos
  et~al.}(2012)\citenamefont{Jim{\'e}nez-Hoyos, Henderson, Tsuchimochi, and
  Scuseria}}]{jimenez2012projected}
\bibinfo{author}{\bibfnamefont{C.~A.} \bibnamefont{Jim{\'e}nez-Hoyos}},
  \bibinfo{author}{\bibfnamefont{T.~M.} \bibnamefont{Henderson}},
  \bibinfo{author}{\bibfnamefont{T.}~\bibnamefont{Tsuchimochi}},
  \bibnamefont{and} \bibinfo{author}{\bibfnamefont{G.~E.}
  \bibnamefont{Scuseria}}, \bibinfo{journal}{J. Chem. Phys.}
  \textbf{\bibinfo{volume}{136}}, \bibinfo{pages}{164109}
  (\bibinfo{year}{2012}).

\bibitem[{\citenamefont{Stuber and Paldus}(2003)}]{stuber2003symmetry}
\bibinfo{author}{\bibfnamefont{J.}~\bibnamefont{Stuber}} \bibnamefont{and}
  \bibinfo{author}{\bibfnamefont{J.}~\bibnamefont{Paldus}},
  \bibinfo{journal}{Fundamental World of Quantum Chemistry, A Tribute Volume to
  the Memory of Per-Olov L{\"o}wdin} \textbf{\bibinfo{volume}{1}},
  \bibinfo{pages}{67} (\bibinfo{year}{2003}).

\bibitem[{\citenamefont{Li and Paldus}(2009{\natexlab{a}})}]{li2009symmetry}
\bibinfo{author}{\bibfnamefont{X.}~\bibnamefont{Li}} \bibnamefont{and}
  \bibinfo{author}{\bibfnamefont{J.}~\bibnamefont{Paldus}},
  \bibinfo{journal}{Phys. Chem. Chem. Phys.} \textbf{\bibinfo{volume}{11}},
  \bibinfo{pages}{5281} (\bibinfo{year}{2009}{\natexlab{a}}).

\bibitem[{\citenamefont{Li and Paldus}(2009{\natexlab{b}})}]{li2009symmetry2}
\bibinfo{author}{\bibfnamefont{X.}~\bibnamefont{Li}} \bibnamefont{and}
  \bibinfo{author}{\bibfnamefont{J.}~\bibnamefont{Paldus}},
  \bibinfo{journal}{J. Chem. Phys.} \textbf{\bibinfo{volume}{130}},
  \bibinfo{pages}{084110} (\bibinfo{year}{2009}{\natexlab{b}}).

\bibitem[{\citenamefont{Colpa}(1986{\natexlab{a}})}]{colpa1986diagonalization}
\bibinfo{author}{\bibfnamefont{J.}~\bibnamefont{Colpa}},
  \bibinfo{journal}{Physica A} \textbf{\bibinfo{volume}{134}},
  \bibinfo{pages}{377} (\bibinfo{year}{1986}{\natexlab{a}}).

\bibitem[{\citenamefont{Colpa}(1986{\natexlab{b}})}]{colpa1986diagonalization2}
\bibinfo{author}{\bibfnamefont{J.}~\bibnamefont{Colpa}},
  \bibinfo{journal}{Physica A} \textbf{\bibinfo{volume}{134}},
  \bibinfo{pages}{417} (\bibinfo{year}{1986}{\natexlab{b}}).

\bibitem[{\citenamefont{Fukutome}(1981)}]{fukutome1981unrestricted}
\bibinfo{author}{\bibfnamefont{H.}~\bibnamefont{Fukutome}},
  \bibinfo{journal}{Internat. J. Quantum Chem.} \textbf{\bibinfo{volume}{20}},
  \bibinfo{pages}{955} (\bibinfo{year}{1981}).

\bibitem[{\citenamefont{Wigner}(1960)}]{wigner1960normal}
\bibinfo{author}{\bibfnamefont{E.~P.} \bibnamefont{Wigner}},
  \bibinfo{journal}{J. Math. Phys.} \textbf{\bibinfo{volume}{1}},
  \bibinfo{pages}{409} (\bibinfo{year}{1960}).

\bibitem[{\citenamefont{Ring and Schuck}(2005)}]{ring2005nuclear}
\bibinfo{author}{\bibfnamefont{P.}~\bibnamefont{Ring}} \bibnamefont{and}
  \bibinfo{author}{\bibfnamefont{P.}~\bibnamefont{Schuck}},
  \emph{\bibinfo{title}{The nuclear many-body problem}}
  (\bibinfo{publisher}{Springer}, \bibinfo{address}{Berlin},
  \bibinfo{year}{2005}).

\bibitem[{\citenamefont{Egido and Ring}(1980)}]{Egido}
\bibinfo{author}{\bibfnamefont{J.~L.} \bibnamefont{Egido}} \bibnamefont{and}
  \bibinfo{author}{\bibfnamefont{P.}~\bibnamefont{Ring}},
  \bibinfo{journal}{Phys. Lett. B} \textbf{\bibinfo{volume}{95}},
  \bibinfo{pages}{331} (\bibinfo{year}{1980}).

\bibitem[{\citenamefont{Tohyama and Schuck}(2004)}]{toyhama}
\bibinfo{author}{\bibfnamefont{M.}~\bibnamefont{Tohyama}} \bibnamefont{and}
  \bibinfo{author}{\bibfnamefont{P.}~\bibnamefont{Schuck}},
  \bibinfo{journal}{Eur. Phys. J. A} \textbf{\bibinfo{volume}{19}},
  \bibinfo{pages}{203} (\bibinfo{year}{2004}).

\bibitem[{\citenamefont{Watanabe and Murayama}(2012)}]{watanabe2012unified}
\bibinfo{author}{\bibfnamefont{H.}~\bibnamefont{Watanabe}} \bibnamefont{and}
  \bibinfo{author}{\bibfnamefont{H.}~\bibnamefont{Murayama}},
  \bibinfo{journal}{Phys. Rev. Lett.} \textbf{\bibinfo{volume}{108}},
  \bibinfo{pages}{251602} (\bibinfo{year}{2012}).

\bibitem[{\citenamefont{Hidaka}(2013)}]{Hidaka}
\bibinfo{author}{\bibfnamefont{Y.}~\bibnamefont{Hidaka}},
  \bibinfo{journal}{Phys. Rev. Lett.} \textbf{\bibinfo{volume}{110}},
  \bibinfo{pages}{091601} (\bibinfo{year}{2013}).

\bibitem[{\citenamefont{Friedman and Wilets}(1970)}]{PhysRevC.2.892}
\bibinfo{author}{\bibfnamefont{W.~A.} \bibnamefont{Friedman}} \bibnamefont{and}
  \bibinfo{author}{\bibfnamefont{L.}~\bibnamefont{Wilets}},
  \bibinfo{journal}{Phys. Rev. C} \textbf{\bibinfo{volume}{2}},
  \bibinfo{pages}{892} (\bibinfo{year}{1970}).

\bibitem[{\citenamefont{Rodr\'iguez-Guzm\'an
  et~al.}(2000)\citenamefont{Rodr\'iguez-Guzm\'an, Egido, and
  Robledo}}]{PhysRevC.62.054308}
\bibinfo{author}{\bibfnamefont{R.~R.} \bibnamefont{Rodr\'iguez-Guzm\'an}},
  \bibinfo{author}{\bibfnamefont{J.~L.} \bibnamefont{Egido}}, \bibnamefont{and}
  \bibinfo{author}{\bibfnamefont{L.~M.} \bibnamefont{Robledo}},
  \bibinfo{journal}{Phys. Rev. C} \textbf{\bibinfo{volume}{62}},
  \bibinfo{pages}{054308} (\bibinfo{year}{2000}).

\bibitem[{Foo()}]{Footnote1}
\bibinfo{howpublished}{Note that if $\mu \to \infty$, then $\mathbf{M} \,
  \mathbf{Q} = \bm{0}$, and one could identify $\mathbf{Q}$ as a proper mode
  satisfying the normalization conditions stated before. In the $\mu \to
  \infty$ limit, the kinetic energy associated with the mode vanishes, and one
  may expect an approximate symmetry restoration scheme to yield no additional
  correlation.}

\bibitem[{\citenamefont{Frisch et~al.}()\citenamefont{Frisch, Trucks, Schlegel,
  Scuseria, Robb, Cheeseman, Scalmani, Barone, Mennucci, Petersson
  et~al.}}]{g09}
\bibinfo{author}{\bibfnamefont{M.~J.} \bibnamefont{Frisch}},
  \bibinfo{author}{\bibfnamefont{G.~W.} \bibnamefont{Trucks}},
  \bibinfo{author}{\bibfnamefont{H.~B.} \bibnamefont{Schlegel}},
  \bibinfo{author}{\bibfnamefont{G.~E.} \bibnamefont{Scuseria}},
  \bibinfo{author}{\bibfnamefont{M.~A.} \bibnamefont{Robb}},
  \bibinfo{author}{\bibfnamefont{J.~R.} \bibnamefont{Cheeseman}},
  \bibinfo{author}{\bibfnamefont{G.}~\bibnamefont{Scalmani}},
  \bibinfo{author}{\bibfnamefont{V.}~\bibnamefont{Barone}},
  \bibinfo{author}{\bibfnamefont{B.}~\bibnamefont{Mennucci}},
  \bibinfo{author}{\bibfnamefont{G.~A.} \bibnamefont{Petersson}},
  \bibnamefont{et~al.}, \emph{\bibinfo{title}{Gaussian~09 {R}evision {H}.21}},
  \bibinfo{note}{{G}aussian {I}nc. {W}allingford {CT }2009}.

\bibitem[{\citenamefont{Dunning~Jr}(1989)}]{dunning1989gaussian}
\bibinfo{author}{\bibfnamefont{T.~H.} \bibnamefont{Dunning~Jr}},
  \bibinfo{journal}{J. Chem. Phys.} \textbf{\bibinfo{volume}{90}},
  \bibinfo{pages}{1007} (\bibinfo{year}{1989}).

\bibitem[{\citenamefont{Coulson and Fischer}(1949)}]{prof1949xxxiv}
\bibinfo{author}{\bibfnamefont{P.~C.} \bibnamefont{Coulson}} \bibnamefont{and}
  \bibinfo{author}{\bibfnamefont{M.~I.} \bibnamefont{Fischer}},
  \bibinfo{journal}{Philos. Mag.} \textbf{\bibinfo{volume}{40}},
  \bibinfo{pages}{386} (\bibinfo{year}{1949}).

\bibitem[{\citenamefont{L{\"o}wdin and Mayer}(1992)}]{lowdin1992some}
\bibinfo{author}{\bibfnamefont{P.-O.} \bibnamefont{L{\"o}wdin}}
  \bibnamefont{and} \bibinfo{author}{\bibfnamefont{I.}~\bibnamefont{Mayer}},
  \bibinfo{journal}{Adv. Quantum Chem.} \textbf{\bibinfo{volume}{24}},
  \bibinfo{pages}{79} (\bibinfo{year}{1992}).

\bibitem[{\citenamefont{Jiménez-Hoyos
  et~al.}(2011)\citenamefont{Jiménez-Hoyos, Henderson, and Scuseria}}]{ghf}
\bibinfo{author}{\bibfnamefont{C.~A.} \bibnamefont{Jiménez-Hoyos}},
  \bibinfo{author}{\bibfnamefont{T.~M.} \bibnamefont{Henderson}},
  \bibnamefont{and} \bibinfo{author}{\bibfnamefont{G.~E.}
  \bibnamefont{Scuseria}}, \bibinfo{journal}{J. Chem. Theory Comput.}
  \textbf{\bibinfo{volume}{7}}, \bibinfo{pages}{2667} (\bibinfo{year}{2011}).

\bibitem[{\citenamefont{Weinberg}(1972)}]{Weinberg}
\bibinfo{author}{\bibfnamefont{S.}~\bibnamefont{Weinberg}},
  \bibinfo{journal}{Phys. Rev. Lett.} \textbf{\bibinfo{volume}{29}},
  \bibinfo{pages}{1698} (\bibinfo{year}{1972}).

\end{thebibliography}
\end{document}